# Graphical Abstract

## Video traffic identification with novel feature extraction and selection method

Licheng Zhang,Shuaili Liu,Qingsheng Yang,Zhongfeng Qu,Lizhi Peng

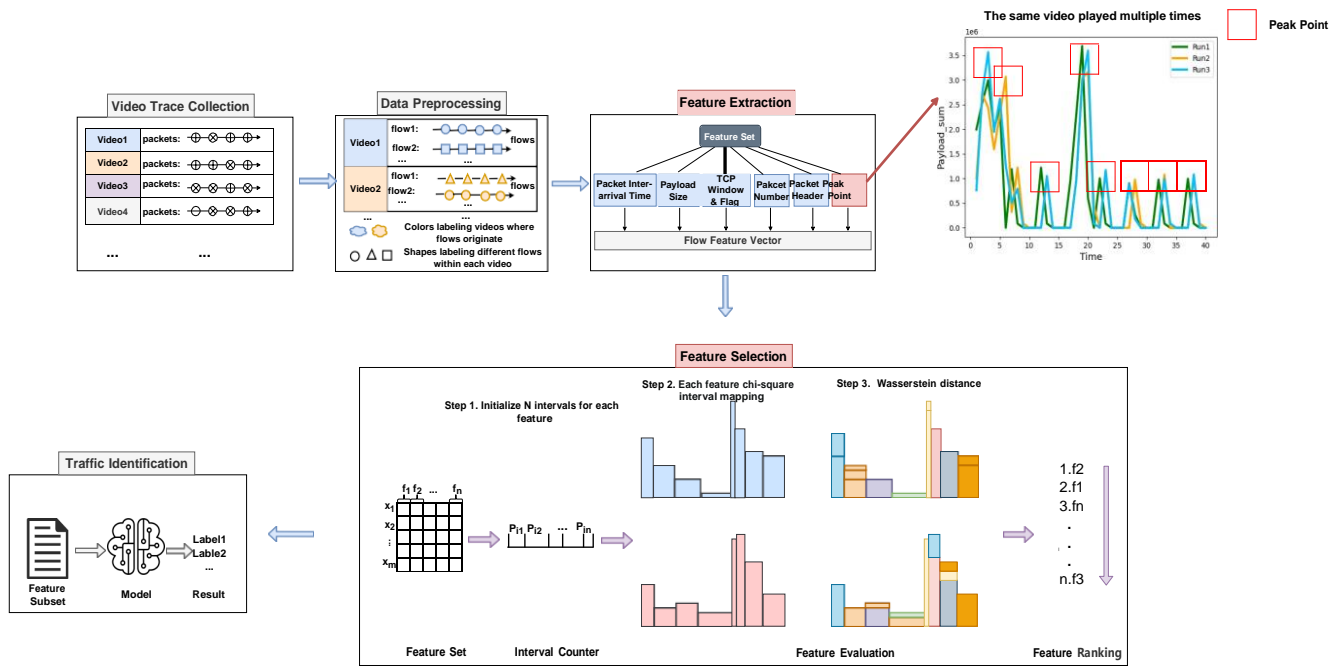

# Highlights

**Video traffic identification with novel feature extraction and selection method**

Licheng Zhang,Shuaili Liu,Qingsheng Yang,Zhongfeng Qu,Lizhi Peng

- A new feature extraction method based on video traffic peak point is proposed.
- A novel adaptive distribution distance-based feature selection (ADDFS) method is proposed.
- Video scene traffic and cloud gaming video traffic are collected for researches and machine learning models training.

# Video traffic identification with novel feature extraction and selection method

Licheng Zhang[a,1], Shuaili Liu[a,1], Qingsheng Yang[a], Zhongfeng Qu[b] and Lizhi Peng[a,c,*]

[a]*Shandong Provincial Key Laboratory of Network Based Intelligent Computing, University of Jinan, Jinan, 250022, China*
[b]*School of Mathematical Sciences, University of Jinan, Jinan, 250022, China*
[c]*Quancheng Laboratory, Jinan, 250022, China*

ABSTRACT

In recent years, the rapid rise of video applications has led to an explosion of Internet video traffic, thereby posing severe challenges to network management. Therefore, effectively identifying and managing video traffic has become an urgent problem to be solved. However, the existing video traffic feature extraction methods mainly target at the traditional packet and flow level features, and the video traffic identification accuracy is low. Additionally, the issue of high data dimension often exists in video traffic identification, requiring an effective approach to select the most relevant features to complete the identification task. Although numerous studies have used feature selection to achieve improved identification performance, no feature selection research has focused on measuring feature distributions that do not overlap or have a small overlap. First, this study proposes to extract video-related features to construct a large-scale feature set to identify video traffic. Second, to reduce the cost of video traffic identification and select an effective feature subset, the current research proposes an adaptive distribution distance-based feature selection (ADDFS) method, which uses Wasserstein distance to measure the distance between feature distributions. To test the effectiveness of the proposal, we collected a set of video traffic from different platforms in a campus network environment and conducted a set of experiments using these data sets. Experimental results suggest that the proposed method can achieve high identification performance for video scene traffic and cloud game video traffic identification. Lastly, a comparison of ADDFS with other feature selection methods shows that ADDFS is a practical feature selection technique not only for video traffic identification, but also for general classification tasks.

## 1. Introduction

At present, most Internet users watch videos daily, resulting in a rapid increase in video traffic. According to Ericsson's mobility report, video traffic will account for 80 % of all mobile data traffic by 2028[1]. Moreover, cloud gaming users currently account for 61.94 % of all Internet users, and market revenue from cloud gaming is expected to triple by 2025 [2]. However, the explosive video traffic has placed a heavy burden on network management departments and network operators, and poses new challenges to existing network infrastructure and protocols. Therefore, effective identification and management of video traffic, particularly game video traffic, have become an important research topic for network management.

Some researchers have explored video traffic identification over the past decade. In the early research of computer vision, video content identification and classification is an important research topic. In general, some features related to video content, such as image shapes, textures, and video text, are extracted to identify video contents[3]. However, this video content identification method is not applicable from the network traffic perspective. The rapid progress of network traffic identification provides a possible solution to this problem. Given that the video transmission mechanism will cause the video content leakage, most existing researchers have extracted traffic features related to video transmission, such as flow statistics feature[4][5], application data unit[6], bit per second[7], burst[8], etc, and used them thereafter to complete the prediction of video QoS and QoE, identification of video application type and video title. Unfortunately, they did not focus on identifying video scene traffic. Additionally, cloud game, as an emerging game mode, is essentially a way of video flow transmission, which is potentially harmful to teenagers. As far as we know, there is no research work reported about identifying cloud game traffic. Modeling and analysis are the main points of the research about cloud game traffic[9][10]. Some researchers have also begun to focus on improving the quality of experience and service of cloud game traffic[11][12], but not identification of cloud gaming video from the traffic level. Therefore, to extract effective features and then build machine learning models for video identification becomes an urgent concern.

A key issue is that from the perspective of feature extraction, current research studies mainly focus on traditional features such as packet-level, flow-level, and burst. These features can not achieve the ideal video identification effect, and further research on video scene traffic feature extraction is needed. Besides, another key problem that should be further discussed is that the quality of extracted or selected features can significantly and directly impact the performance of identification. Irrelevant or redundant features can cause unnecessary cost and time overhead, even negative impact for the model identification. Therefore, a well-performing feature selection method is crucial for traffic identification.

*Corresponding author

plz@ujn.edu.cn (L. Peng)

[1]Licheng Zhang and Shuaili Liu contributed equally to this paper.

Feature selection methods have three categories: filter, wrapper, and embedded methods. The filter method scores each feature according to feature correlation and sets thresholds to select features, which is independent of the subsequent identification models. Accordingly, filter becomes the most popular feature selection technique, and this study focuses on this type. Filters often use correlation coefficient or mutual information to measure the importance of features. Some researchers have also proposed to use distribution divergence for feature evaluations. Kullback-Leibler (KL) divergence and Jensen-Shannon (JS) divergence are used to measure the similarity between feature distributions. Nevertheless, when the support sets of the two feature distributions do not overlay or have a small overlap, neither of the two evaluation criteria can reflect the similarity of the two feature distributions. To the best of our knowledge, no relevant research has been conducted to solve this feature selection problem.

To solve the preceding problems, we make the following contributions.

- A new feature extraction method based on video traffic peak point is proposed, which can be used as an effective supplement of traditional packet and flow level features.

- A novel adaptive distribution distance-based feature selection (ADDFS) method is likewise proposed. We construct a relatively comprehensive video traffic feature set and use ADDFS to select an effective feature subset to achieve the best identification effect.

- We design a video traffic collection architecture to collect different video traffic data, mainly including video scene traffic and cloud gaming video traffic.

- To verify the effectiveness of the extracted feature set and the proposed ADDFS, a set of experiments are conducted on the collected data. We apply six learning models for video traffic identification. Video scene and game video traffic identification accuracies are above 92% and 99%, respectively. In addition, the comparison of ADDFS with other feature selection methods suggests that ADDFS has superior performance in video traffic identification.

Roadmap: Sec. 2 reviews the related research on video traffic identification and feature selection. In Sec. 3, introduces the video traffic identification method. Sec. 4 presents the performance measures and experimental results. Lastly, Section 5 concludes this paper in Sec. 5.

## 2. Related research
### 2.1. Video traffic identification

Three kinds of traffic identification methods have been used for video traffic identification: port-based, deep packet inspection, and machine learning algorithms. The first two methods have become ineffective owing to the dynamic port and encryption techniques, making machine learning-based method a widely used technology in traffic identification[13][14]. Researchers have made some preliminary exploration of video traffic identification in the last decade. In 2012, Ameigeiras et al.[15] analyzed YouTube's video traffic generation pattern to predict the quality of video watching experience. Given that early YouTube videos were based on Flash, which is no longer used, this method is no longer effective for current video traffic. Wampler et al.[16] analyzed encrypted IP video traffic according to different video encoding modes and found information leakage of the encrypted IP traffic. In 2015, Shi et al.[4] used packet size distribution to study the impact of protocol and encoding on identifying the source of video streams. However, these studies have not addressed video content identification. Reed et al.[7] proposed a new bit per peak feature extraction method, and used these features for classifying video stream titles. To provide better QoS for video services, Dong et al.[17] proposed a fine-grained classifier for video traffic, which uses flow statistical features to analyze whether or not the targeted traffic is a video flow. Mangla et al.[18] used packet headers from video traffic to estimate the video quality of experience. Wu et al.[6] proposed a method to extract application data unit combination from QUIC video flows, and used it as a fingerprint to identify video traffic. However, this method does not work with HTTPS.

At present, only a few studies focus on cloud gaming video traffic identification, and the existing research has mainly concentrated on the analysis and modeling of cloud gaming traffic and improving the cloud gaming experience. To detect cloud gaming video traffic, Manzano et al.[19] attempted to find the general attributes of cloud gaming traffic by analyzing the traffic of different game types under the same cloud gaming platform. To improve the experience of end game users, Suznjevic et al.[20] collected cloud gaming video samples to calculate video indicators from the time and space dimensions. Thereafter, they analyzed the relationship among game types, cloud gaming video traffic features, and video indicators. In 2015, Amiri et al.[21] proposed a paradigm for software defined network (SDN) controller to reduce cloud gaming delay. Carrascosa et al.[22] deeply analyzed the traffic characteristics of the different protocols, traffic generation patterns, packet size, and packet time probability distributions of the stadia platform. Thereafter, they further compared the differences between different game video streams on this platform. All these studies have focused minimally on identifying video scene traffic and cloud gaming video traffic, which is the focus of this paper.

### 2.2. Feature selection

Over the years, three types of methods have been developed for feature selection: filter, wrapper, and embedded methods. The filter feature selection method has the advantage of high efficiency. These methods depend on specific evaluation criteria to evaluate the correlation between features and categories and select target features. Zhang et al.[23] used KL divergence to analyze the correlation and redundancy of different class labels. Mousselly et al.[24] proposed an adapted JS divergence to measure the probability distribution of related tags, which can effectively deal with the fluctuation of feature samples. Nevertheless, their research does not consider the measurement problems of small overlap and no overlap between feature distributions. The wrapper method uses the classifier's performance as evaluation criterion to select a subset of features with the strongest discrimination ability by using a given classifier. Compared with filter feature selection, a wrapper can often hit markedly high accuracies, but it is time consuming. For the embedded method, the feature selection process is integrated with the training process of the classification model. This type of method can solve the excessive redundancy problem of filters and excessive time requirement of wrappers. Therefore, the embedded feature selection method is a compromise between the filter and wrapper methods.

Feature selection is vital for traffic identification because all types of features are extracted from raw traffic data. Many of these features are redundant or with no contribution for identification. Therefore, researchers have attempted to develop effective methods to evaluate and select traffic features in recent years. In 2018, Su et al.[25] proposed a method of learning automata to select significant features in the KDD CUP 99 data set. Selvakumar et al.[26] worked on the same data set and combined the filter and wrapper-based methods with the firefly algorithm to obtain an effective feature set. Fahad et al.[27] combined the results of various feature selection methods and proposed an integrated method of network traffic feature selection, which realizes the classification of network application traffic. To solve the class imbalance problem in the network traffic classification, Liu et al.[28] proposed a class-oriented feature selection method, combining global and local metrics and using weighted symmetric uncertainty characteristics to remove redundant features in the feature set. McGaughey et al.[29] proposed a fast orthogonal search method to rapidly select a subset of features with discriminative power from a large set of features.

Some researchers have begun to apply feature selection methods to video traffic identification. In 2017, Dong et al.[30] combined ReliefF and particle swarm optimization (PSO) to solve the excessive dimensionality problem in network traffic classification. In the same year, they also used the coefficient of variation method to select video traffic features and completed video traffic classification[31]. Wu et al.[32] used a linear consistency-constrained method to select features for multimedia traffic classification and completed instance purification in the selection process. To improve the quality of experience (QoE) of YouTube videos, Seufert et al.[33] selected features on the basis of the correlation between features and tested the prediction performance under different feature subsets. Yang et al.[34] used a consistency-based feature selection method to select the QoE value of videos, Thereafter they used their statistical features for feature selection to complete the fine-grained classification of online videos. To the best of our knowledge, no research using has been conducted on distribution distance to measure the similarity between video traffic feature distributions. Therefore, this paper overcomes this drawback, by using Wasserstein distance to adaptively measure the similarity between feature distributions, and build an effective feature selection algorithm thereafter.

## 3. Methodology

Fig. 1 shows the framework of the proposed video traffic identification method. First, we automatically obtained the video traffic data in a controlled environment and preprocessed the collected raw traffic data. Second, feature extraction was carried out on the preprocessed data to form a large-scale feature set. Finally, to obtain a good identification effect, we propose a feature selection method based on adaptive distribution distance.

### 3.1. Data collection

Only a few public video traffic data sets are available for video traffic identification research, and not mention video scene traffic data sets. Therefore, we collected a video scene traffic data set (VS-UJN-2022) and cloud gaming video traffic data set (CG-UJN-2022) in a controlled campus environment. Our collection client computer is equipped with Intel Core i5-7500 CPUs and runs on Windows 10. The entire collection process was completed in approximately five months in the year of 2022.

**Video Scene Traffic Data Collection.** Video scene traffic data we collected can be divided into two main categories: action and static scene videos. Action scene videos are from science fiction action films, such as *Transformers*, *The Avengers*, *Iron Man*, *Fast & Furious*, and *Pirates*. Unlike action scene videos, static scene videos have a relatively simple scene, such as class scenes, natural scenery, tourism documentary, and light music video. We collected both types of data from YouTube and Bilibili (a widely used Chinese video cite). Fig. 2 shows the capturing architecture of video scene traffic.

We downloaded the videos of the aforementioned video categories completely to a client computer and used FFmpeg thereafter to cut the original video into video clips with fixed duration of 120 s. We regard a 120 s video segment as a scene because such a segment can provide sufficient network features for coarse-grained video scene identification. For all action movie segments, we manually selected the segment with actions to form the action scene video set. The static video has few scenes, so no filtering was performed.

By combining the Python installed Selenium library with the Xpath Helper plug-in in Google Chrome, we automatically uploaded the fixed video segment to YouTube and Bilibili. We used Selenium and T-shark, and accomplished the automatic on demand of the target video and the video traffic is automatically collected when a video is playing. When the client's computer was playing a video, all other network applications were closed to avoid generating noise traffic. Lastly, videos on demand were saved in our accounts, preventing the generation of advertising video traffic.

Video traffic identification with novel feature extraction and selection method

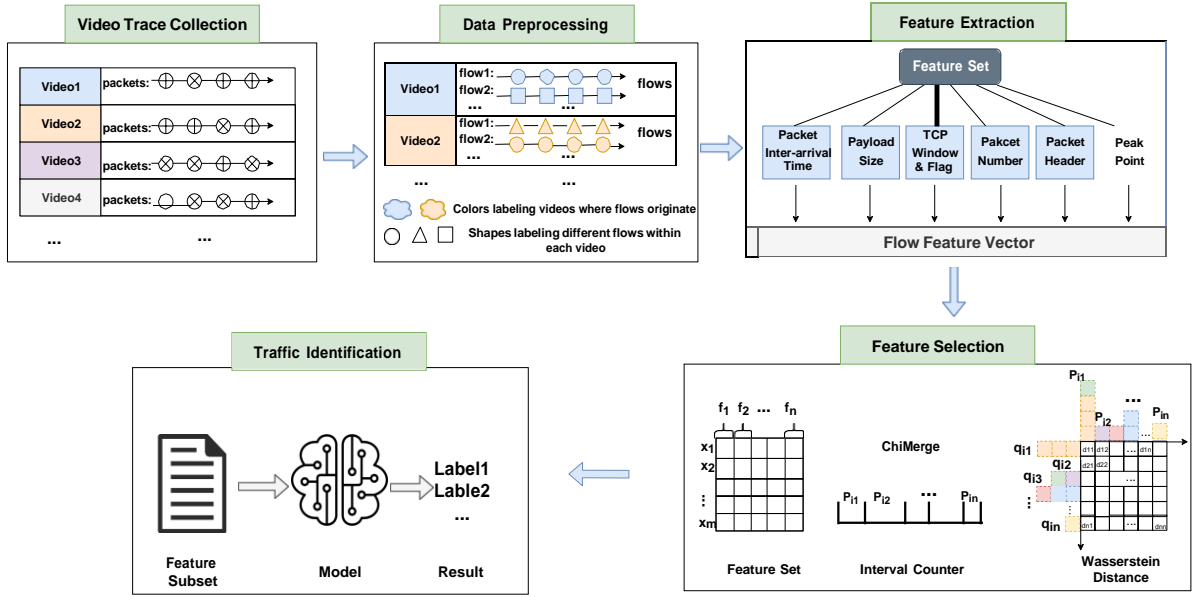

**Fig. 1:** The framework of the proposed video traffic identification method.

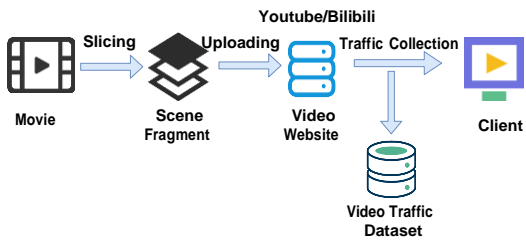

**Fig. 2:** The architecture for video scene traffic collection.

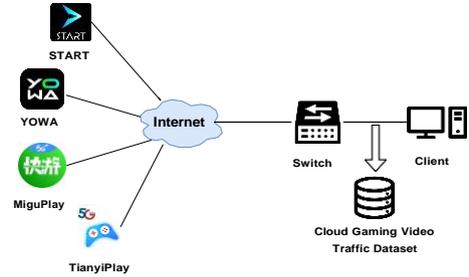

**Fig. 3:** The architecture of cloud gaming video traffic collection.

**Cloud Gaming Video Traffic Data Collection.** The cloud gaming platform we visited included Tencent Start, YOWA cloud gaming, MiguPlay, and Tianyi cloud gaming. We deployed Wireshark on the client computer to collect raw cloud gaming video traffic data. The collection architecture is shown in Fig. 3. To ensure the comparability of some extracted features, we set Wireshark to automatically save the collected data as a .pcap file every 2 min. We shut down other applications while collecting targeted traffic to prevent non-targeted traffic from being generated. We also captured parts of the background traffic, mainly covering the most popular application types, such as non-game video, online meetings, and chatting. All traffic data were collected on the same network environment using the same collection device. Tables 1 and 2 show the details of the collected traffic data.

### 3.2. Data preprocessing

Raw traffic data were captured in .pcap file format, and the consequent data preprocessing includes a set of operations that convert these raw data into a processable format as required.

First, by using PcapPlusPlus, a C++ library installed on Linux, the collected traffic data are grouped into flows according to the five-tuple information: {src IP, src port, dst IP, dst port, protocol (TCP/UDP)}. Given that YOWA, MiguPlay, and TianyiPlay use UDP as the transport layer protocol, we focus on UDP packets when analyzing the three platforms and TCP packets for the rest of the traffic.

**Table 1**
The details of video traffic data

| Dataset | Platform | Flows | Bytes |
|---|---|---|---|
| VS-UJN-2022 | YouTube | Action-821 | 4,994,062,797 |
| | | Static-538 | 3,294,957,584 |
| | Bilibili | Action-789 | 5,956,426,782 |
| | | Static-767 | 7,672,791,660 |
| CG-UJN-2022 | START | 83 | 3,401,167,754 |
| | YOWA | 58 | 2,116,354,048 |
| | MiguPlay | 61 | 2,548,042,936 |
| | TianyiPlay | 63 | 3,132,444,936 |

Second, elephant flows are selected from the mice flows. Elephant flow is a continuous flow that transmits a large amount of data over a network link, whereas mice flow is a short flow that contains a few packets. Mice flows often contain minimal useful information, and we only focus on elephant flows because video traffic are all elephant flows. We eliminate the mice flows according to the number of non-zero payloads. According to experience, the number of non-zero payloads was set to 500 in this paper and those flows with under 500 packets are considered mice flows to be eliminated.

**Table 2**
Background traffic data

|  | Flows | Bytes |
|---|---|---|
| PC Video | 42 | 704,530,546 |
| PC Live Steaming | 40 | 397,123,864 |
| Web Browsing | 57 | 73,440,492 |
| Online Meeting | 5 | 21,501,967 |
| Chatting | 56 | 134,386,569 |
| Online Shopping | 42 | 142,277,293 |
| File Download | 40 | 57,409,424 |

Lastly, the server name indication extension field in the Client Hello packet is used to determine whether or not the captured flow is the target flow. For example, if the server name indication contains the "*.googlevideos.com" string, then this flow will be considered a flow from YouTube.

### 3.3. Feature extraction

In most real cases, raw data cannot be directly used for building learning model because of data conflict, duplication, lack of classification, and so on. Therefore, formatted feature data are extracted from the raw data to represent specific problems. This process is called feature extraction, which is a critical machine learning step, that effectively reduces the volume of raw data. Feature extraction is also the basic and important step for the consequent learning model building.

In this study, 89 statistical features are extracted from preprocessed data for video traffic, some of which are traditional traffic features, while others are the proposed peak point features. For each flow, we analyze the characteristics of the packet sequence from three directions: upstream, downstream, and all packets (both directions). Upstream data are data sent from local clients to the Internet, whereas downstream data are those transmitted from the Internet to local clients. The following types of traditional traffic features we extracted are:

**Packet inter-arrival time**. According to the arrival time of packets in each flow, the time difference between the two packets is the packet inter-arrival time.

**Payload size**. Payload size is the one-time transmission data size of the TCP message. Payload size is related to the amount of data provided by the application and buffer. However, payload is often below 1460, which may vary with the configuration of MTU.

**TCP window size**. To obtain the optimal connection rate, the TCP window is used to control the flow rate. Such features are used to inform the sender of the amount of data it can receive, thereby achieving flow control.

**TCP flag**. The TCP flag value represents the purpose of the current client request. The eight flags are as follows: FIN, SYN, PSH, ACK, RST, URG, ECE, and CWR. For PSH and URG, we separately count the numbers of upstream and downstream.

**Packet number**. Total number of packets, ratio of packets downstream to those upstream, and average number of packets per second.

**Packet header**. Total length of the packet header in three directions and ratio of the total length of the packet header to the sum of the packet length in each direction.

For the first three features, the statistics, including mean value, standard deviation (std), maximum, and minimum, are calculated for up and down directions and for dual directions. Moreover, total size of the TCP window is calculated as a feature. Different videos have different styles, and the corresponding traffic characteristics also have different patterns. For example, action scene videos are relatively intense. Hence, data packets are sent more frequently within a period, while static scene videos generate relatively few data packets. Thus, we define maximum data transmission within a period as the peak point, to extract the characteristics of the packet payload and byte rate peak points for video traffic identification.

**Payload peak point**. Assume there are $d$ packets in a flow and the sequence of the packets is $Pkt_1, Pkt_2, ..., Pkt_d$. Payload size of the $s$th packet is denoted as $pay_s$. If $pay_s \geq pay_{s-1}$ and $pay_s \geq pay_{s+1}$ ($1 < s \leq d-1$), then payload reaches a peak in a certain period of time, which is defined as the payload peak point (PPP). We use a set of counters $c^1, c^2, ..., c^l$ to count the number of PPP every $\alpha$ s in the first $\beta$ s of the $t$th flow, then the $\theta$ is calculated as follows:

$$\theta = \frac{\beta}{\alpha}. \quad (1)$$

where $\alpha$ and $\beta$ are set as 5 and 60, respectively, in this paper. Thereafter, go through the whole flow sequence ($f_1, f_2, ..., f_t$) to obtain the count matrix $CT$.

$$CT = \begin{bmatrix} c_1^1 & c_2^1 & \cdots & c_\theta^1 \\ c_1^2 & c_2^2 & \cdots & c_\theta^2 \\ \vdots & \vdots & \ddots & \vdots \\ c_1^t & c_2^t & \cdots & c_\theta^t \end{bmatrix} \quad (2)$$

On the basis of $CT$, the mean and standard deviation of the PPP of the $t$th flow is obtained as follows:

$$M_t = \frac{1}{\theta} \sum_{a=1}^{\theta} c_a^t, \quad (3)$$

$$Std_t = \sqrt{\frac{1}{\theta}[(c_1^t - M_t)^2 + (c_2^t - M_t)^2 + ... + (c_\theta^t - M_t)^2]}. \quad (4)$$

Similarly, we obtain the mean and std of the PPP for all flows. The maximum, minimum, and total number of PPPs in three directions are also extracted. However, PPP alone is insufficient because the change in packet payload is not significant when some scene videos are transmitted stably. Hence, we propose the byte rate peak point (BRPP).

**BRPP**. Assuming that the sum of the packet payloads (SPP) in period $T$ is calculated as follows:

$$SPP = \sum_{b=1}^{H} pay_b, \quad (5)$$

where $H$ is the total number of packets within $T$ s, and $pay_b$ is the size of the $b$th packet payload. Thereafter, byte rate (BR) in this period is defined as follows:

$$BR = \frac{SPP}{T}. \quad (6)$$

Similar to the PPP, if BR satisfies the preceding definition of peak point, then the point is BRPP. In this study, $T$ is set to 1 s, and we count the number of BRRPs from three directions.

**BRPP with sliding windows**. To catch continuous video information more accurately, we design sliding windows to extract the size of peak points as feature vectors based on BRPP. Length of the sliding window is $L$ and offset factor is denoted by $Z$. In this study, $L$ and $Z$ are set to 3 and 0.5, respectively. For a packet sequence $(P\ kt_1, P\ kt_2, ..., P\ kt_d)$, we calculate the sum of packet size under in time window $L$ and use the offset factor thereafter to move the window to calculate the total packet size in turn. The sum of packet size in the $z$th window can be calculated as follows:

$$r_z = \sum_{pt=0}^{L} pktLen_{pt}, \quad (7)$$

where $pt$ is the arrival time of the packet and $pktLen_{pt}$ is the packet size at $pt$th s. The processed sequence $R=(r^{pt}, r, ..., r_n)$ is obtained, where $n$ is the number of sliding windows. If the value in the sequence meets the definition of the preceding peak point, then the point is defined as BRPP with sliding windows (BRPPSW). Therefore, we will obtain the sequence $R\_F =(r_1, r_2, ..., r_u)$ of BRPPSW, which is a subset of $R$. An example of the BRPPSW extraction is shown in Fig. 4.

We calculate the mean, std, maximum and minimum values of BRPPSW from three directions. The first, second, and third quartile of BRPPSW are also extracted as features.

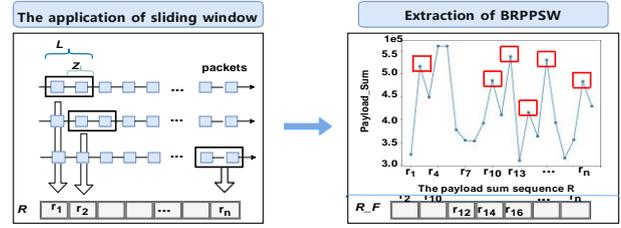

**Fig. 4:** The extraction process of BRPPSW.

### 3.4. Feature selection

By the previous feature extraction, a comprehensive feature set with 89 features is obtained. However, note that we do not consider whether these features are redundant or useless at the extracting process. To select an effective and concise feature subset, we propose an adaptive distribution distance-based feature selection (ADDFS) approach.

Let $X = \{X_1, X_2, ..., X_m\}$ be the data set, $X_i$ ($1 \leqslant i \leqslant$ m) be the $i$th sample data, and $m$ be the number of samples. The feature set is defined as $F = \{f_1, f_2, ..., f_n\}$, where $f_j$ ($1 \leqslant j \leqslant$ n) is the $j$th feature, and $n$ is the number of features, and $x_{ij}$ represents the $j$th feature value of the $i$th sample.

First, Min-Max scaling is used to normalize all feature values of the samples in the dataset to the [0,1] range. The Min-Max scaling formula is as follows:

$$X_{ij} = \frac{X_{ij} - min(X_{.j})}{max(X_{.j}) - min(X_{.j})}, \quad (8)$$

where $max(X_{.j})$ is the maximum value of the $j$th feature, and $min(X_{.j})$ is the minimum value of the $j$th feature.

Second, the supervised ChiMerge algorithm [35] is used to divide each feature into multiple consecutive intervals. For each feature, we first sort all values in ascending order. Thereafter, we group the data with the same feature value into the same interval, and calculate the chi-square value of the interval. Each adjacent chi-square value is calculated and the smallest pair of intervals are merged. This step is repeated until the set maximum binning interval or chi-square stopping threshold is reached. Lastly, the chi-square binning interval of each feature is obtained. The flowchart of the method is shown in Fig. 5. According to empirical values, the maximum binning interval and stop confidence threshold in this paper are set to 15 and 0.95, respectively. The chi-square calculation formula is as follows:

$$\chi^2 = \sum_{\gamma=1}^{G} \sum_{\psi=1}^{C} \frac{(A_{\gamma\psi} - E_{\gamma\psi})^2}{E_{\gamma\psi}}, \quad (9)$$

$$E_{\gamma\psi} = \frac{N_\gamma}{N} \times C_\psi, \quad (10)$$

where $G$ is the number of intervals, $C$ is the number of classes, $A_{\gamma\psi}$ represent the number of samples of the $\psi$th class in the $\gamma$th interval, $E_{\gamma\psi}$ is the expected frequency of $A_{\gamma\psi}$, and $N$, $N_\gamma$, and $C_\psi$ represent the total number of samples, number of samples in the $\gamma$th interval, and number of samples in the $\psi$th class, respectively.

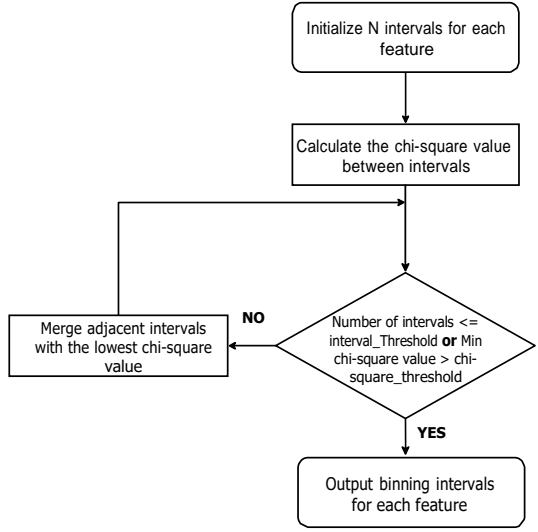

**Fig. 5:** The flowchart of the ChiMerge.

For each feature, the number of samples of a particular feature within the chi-square binning interval in each class is counted. Take the feature $F_j$ as an example. For class $C1$, the distribution of feature $F_j$ on the chi-square binning interval $(p_{11}, p_{12}, ..., p_{1k})$ can be obtained by counting the number of samples of feature $F_j$ in each interval, in which $k$ is the number of chi-square binning intervals for this feature. For class $C2$, the distribution of feature $F_j$ can be calculated as $(p_{21}, p_{22}, ..., p_{2k})$. On this basis, we can obtain the feature distribution matrix $P$ of feature $F_j$ on n classes. In the same manner, the feature distribution of other features on different classes can also be obtained.

$$P_{n \times k} = \begin{bmatrix} p_{11} & p_{12} & \cdots & p_{1k} \\ p_{21} & p_{22} & \cdots & p_{2k} \\ \vdots & \vdots & \ddots & \vdots \\ p_{n1} & p_{n2} & \cdots & p_{nk} \end{bmatrix} \quad (11)$$

Then for each feature, Wasserstein distance is applied to obtain the score. That is, the Wasserstein distance is used to calculate the distribution difference between each two classes. The Wasserstein distance is also called the Earth mover's distance (EMD). The larger the EMD value of the feature in the two classes, the more distinguishable the feature. EMD for each two classes is calculated as follows:

$$W(P_U, P_V) = \inf_{\gamma \sim \Pi(P_U, P_V)} E_{(U,V) \sim \gamma}[\|x - y\|], \quad (12)$$

where $P_U$ and $P_V$ are the feature distribution of a feature on two classes, $\Pi(P_U, P_V)$ is the set of all possible joint distributions $P_U$ and $P_V$, and $W(P_U, P_V)$ represents the mathematical expectation lower bound of $\gamma(x, y)$. EMD for multi-class is calculated as follows:

$$EMD = \sum_{\kappa=1}^{} \sum_{\lambda=\kappa+1}^{} W(P_\kappa, P_\lambda), \quad (13)$$

Lastly, the EMD value for each feature is obtained by calculating the Wasserstein distance, and each feature is sorted using the EMD value in descending order. According to feature ranking, we can select the optimal feature subset as needed. The pseudo-code of ADDFS is shown in Algorithm 1.

---

**Algorithm 1** Feature selection algorithm based on adaptive distribution distance

**Require:** Feature set $F$, classes $C$
**Ensure:** The selected feature subset $S$
1: BEGIN
2: compute $F$ according to equation (8);
3: for each $f \in F$:
4:     Interval=ChiMerge($f$);
5:     // The ChiMerge algorithm is used to divide each feature into multiple consecutive intervals
6:     for each $c \in C$:
7:         Pc=count($f$,c,Interval);
8:         // Calculate the number of samples of the feature in the chi-square binning interval of each class
9:     end for;
10:     for each $p \in P$:
11:         for each $p' \in P$, $p' \neq p$:
12:             compute W($p,p'$) according to equation; (12)
13:             EMD$_f$ += W($p,p'$);
14:         end for;
15:     end for;
16: end for;
17: $S$= sort($EMD$);
18: // All feature is sorted by the EMD value in descending order
19: END;

---

### 3.5. Machine learning model

Six machine learning models are used to identify the extracted and selected features in this study, as shown in Table 3. Noted that we do not focus on the actual machine learning model but on the effect of our proposed method combined with the machine learning model on video traffic identification. In the experimental part, a 10-fold cross-validation is used to train and test the selected learning model. By comparing the identification results of different models, we can choose the model with superior performance for video traffic identification.

## 4. Experiment

### 4.1. Performance measures

The confusion matrix, which is also called error matrix, is an effective method of measuring the performance of the machine learning model. Fig. 6 illustrates the confusion matrix of the binary classification.

**Table 3**
Machine learning models

| Classifier | Abbreviation |
| --- | --- |
| Random Forest | RF |
| Decision Tree | DT |
| Extremely Randomized Trees | ET |
| Multi-Layer Perceptron | MLP |
| K-Nearest Neighbors | KNN |
| Adaptive Boosting | AdaBoost |

| Confusion Matrix | Positive (Predicted) | Negative (Predicted) |
| --- | --- | --- |
| Positive (Real) | True Positive | False Negative |
| Negative (Real) | False Positive | True Negative |

**Fig. 6:** Confusion Matrix for Binary Classification

On the basis of the confusion matrix, accuracy (ACC) and F1 score can be derived as the evaluation criteria in our experiment. ACC in the binary classification task can be defined as follows:

$$ACC = \frac{TP + TN}{TP + FN + TN + FP}, \quad (14)$$

where true positive (TP) and true negative (TN) are the number of positive and negative instances, respectively, correctly predicted by the model. False negative (FN) and false positive (FP) are the number of incorrectly predicted positive and negative instances, respectively, by the model. Precision and recall can be defined as follows:

$$Precision = \frac{TP}{TP + FP}, \quad (15)$$

$$Recall = \frac{TP}{TP + FN}. \quad (16)$$

With precision and recall, F1 score, a widely used performance measure, can be derived as follows:

$$F1 = 2 \times \frac{Precision \times Recall}{Precision + Recall}. \quad (17)$$

### 4.2. Identification model parameter settings

All learning models used in this study have been introduced in Sec. 3.5. Table 4 shows the detailed parameter settings of the selected models. For each model, we use the fixed default parameter settings to ensure fair comparisons.

**Table 4**
Parameter settings of the compared identification models

| Classifier | Parameter |
| --- | --- |
| RF | n_estimator=10, max_depth=None, min_samples_split=2 |
| DT | max_depth=None, min_samples_split=2 |
| ET | n_estimator=10, max_depth=None, min_samples_split=2 |
| MLP | activation=rule, solver=adma |
| KNN | n_estimator=10, metric=euclidean |
| AdaBoost | n_estimator=10 |

### 4.3. Evaluation of ADDFS with video traffic identification

The overall identification performance of video scene traffic and cloud gaming video traffic are first evaluated by using the selected learning models and proposed feature selection algorithms. ADDFS is applied to select 10%, 20%, …, and 90% of the feature set as feature subsets. Thereafter, all selected learning models are used to identify both types of video traffic. The results are presented in Figs. 7 and 8.

From the perspective of the number of selected feature set, for YouTube and Bilibili, the identification effects of most of the learning models hit the optimum at 20% and 60%, respectively, of the feature set and reach a steady state thereafter. For cloud games, the recognition effect of the learning model maintains a small range of fluctuations on different feature subsets. Therefore, ACC and F1 score of most classifiers on different feature subsets reach a relatively stable state. The state is relatively stable when ACC and F1 score fluctuate within a small range as the number of features increases. This situation is particularly evident in YouTube's scene video traffic data. After 20% of the feature set, the identification effect of the machine learning model has reached a stable state.

From the perspective of the learning models, the comparison of ACC and F1 score of different classifiers in the stable state indicates that RF, ET, and AdaBoost outperform the remaining three classifiers. The performance of the decision tree is more stable than that of the other models. Accuracies of RF and AdaBoost on YouTube exceeds 0.95 in the stable state. Moreover, the accuracies of RF, ET, and AdaBoost on Bilibili and cloud gaming are above 0.92 and 0.99, respectively. For other classifiers, Figs. 7(b) show that the identification performance of DT is particularly stable compared with other models. Even though accuracy of MLP shows improvement with an increase in selected feature number regarding the data of YouTube and Bilibili, overall performance is lower than that of the other classification model. KNN also obtained a poor overall performance. As shown in Fig. 8, F1 score also presents a similar performance patterns to that of ACC.

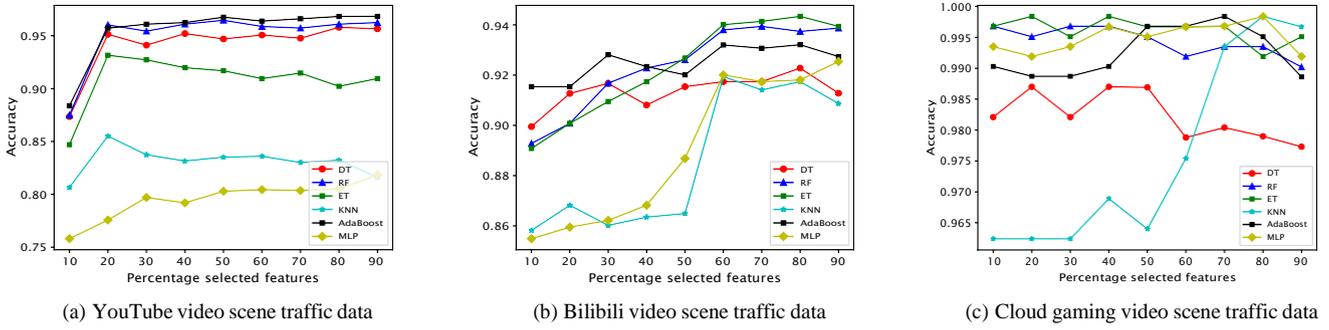

**Fig. 7:** Accuracy results with varying feature number percentage selected by ADDFS

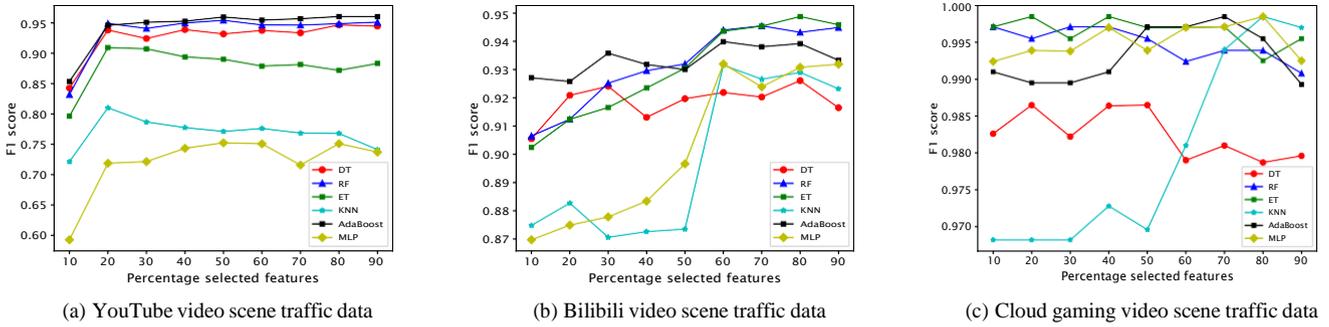

**Fig. 8:** F1 score with varying feature number percentage selected by ADDFS

### 4.4. Evaluation of the effectiveness of peak point features

To test the effectiveness of peak point features, we compare the feature set with and without peak points applying RF as the identification model. Each group of experiments is 10-folder cross-validated, and the mean value regarded is the final result.

The comparison results are shown in Figs. 9(a) and 9(b), in which FS is the complete feature set with peak point features, and FS-PP is the feature set without peak point features. The results of FS are observed to be better than those of FS-PP, particularly for data of video scene traffic on the YouTube platform. That is, ACC and F1 increased by over 3%. For the other two cases, the two evaluation measures also improved slightly with the joining of peak point features. Therefore, the experimental results clearly show that the proposed peak point feature is effective for video traffic identification.

### 4.5. Evaluation of the impact of sliding windows

This subsection evaluates the impact of different sliding window sizes and offset factors on video flow identification on cloud gaming. RF is used as a classifier, and 10-fold cross-validation is again applied.

Fig. 10(a) demonstrates the impact of different sliding window sizes on identification accuracy. Offset factor is set to 0.5. As window size grows, identification accuracy increases initially. Thereafter, it reaches the highest when window size is set to 3. Accuracy decreases thereafter as window size increases. Therefore, we obtain the empirical optimal window size of 3. Fig. 10(b) shows the results with the varying offset value. Note that when offset factor is 0.5, accuracy of video traffic identification hits the highest value. When offset factor increases, recognition accuracy tends to be stable. Thus, we set window size $L$ to 3 and the offset factor $Z$ to 0.5 in our studies.

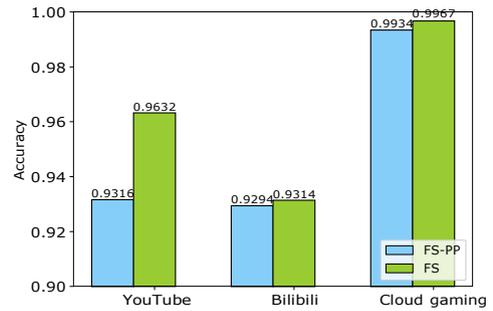

(a) The comparison results of ACC

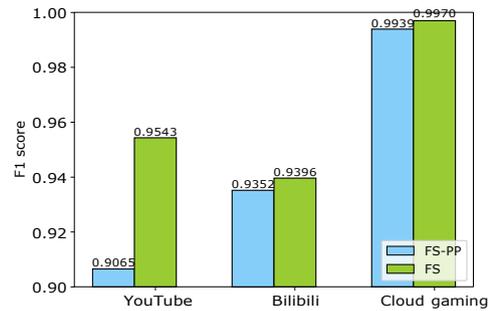

(b) The comparison results of F1 score

**Fig. 9:** The comparison results with/without peak point features

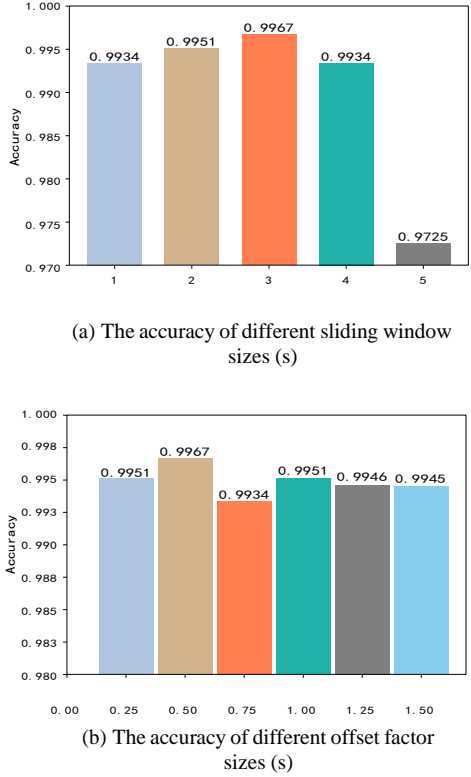

(a) The accuracy of different sliding window sizes (s)

(b) The accuracy of different offset factor sizes (s)

Fig. 10: The impact of sliding window

### 4.6. Evaluation of the ADDFS performance

To further verify the effectiveness of the feature selection algorithm ADDFS, we conduct comparative experiments on 9 public datasets and 3 private traffic datasets (VS-UJN-2022-YouTube, VS-UJN-2022-Bilibili and CG-UJN-2022) with 5 feature selection algorithms. Data sets used in this study are shaown in Table 5. The five compared feature selection methods are Relief[36], Person[37], RFS[38], DDFS[39], and F-score[40]. We use DT as classifier, and accuracies of the compared methods are compared through 10-fold cross-validation. The classification accuracy results are shown in Fig. 11.

As shown in Fig. 11, all compared methods will receive increasing accuracy as the number of selected features increases for most data sets, and reach a relatively steady state thereafter. When the number of selected features is small, ADDFS shows high accuracy compared with the other methods. Note that it has consistently maintained efficient and stable performances for the cases of the optdigits, thyroid, Mushroom, Spambase, soybean, and VS-UJN-2022-Bilibili datasets. Although there are numerous redundant and irrelevant features in the CG-UJN-2022 dataset, ADDFS can still obtain a relatively stable classification accuracy in the early stage.

**Table 5**
Public classification data sets used in the experiments

| Data sets | Instances | Features | Classes | Source |
| --- | --- | --- | --- | --- |
| penbased | 10992 | 16 | 10 | KEEL |
| thyroid | 7200 | 21 | 3 | KEEL |
| optdigits | 5620 | 64 | 10 | KEEL |
| segment | 2310 | 19 | 7 | KEEL |
| wine | 178 | 13 | 3 | KEEL |
| Mushroom | 8124 | 22 | 2 | UCI |
| Spambase | 4601 | 57 | 2 | UCI |
| QSAR biodegradation | 1055 | 41 | 2 | UCI |
| Soybean | 307 | 35 | 19 | UCI |

**Table 6**
Wilcoxon signed rank test of ADDFS versus the compared methods-10% of feature set

| | Accuracy | | |
| --- | --- | --- | --- |
| ADDFS vs. | $R^+$ | $R^-$ | P–value |
| Relief | 77.0 | 1 | 0.0025 |
| Person | 72.0 | 6.0 | 0.0085 |
| RFS | 76.0 | 2.0 | 0.0033 |
| DDFS | 78.0 | 0.0 | 0.0019 |
| F-score | 70.0 | 8.0 | 0.0135 |

To demonstrate that ADDFS has superior performance with a few features, we use Wilcoxon signed-Rank test [41] to verify the performance of different methods with feature sets of 10%, 20%, and 30%. The results are shown in Tables 6, 7, and 8, respectively. In particular, $R^+$ represents the sum of permutation ordinals of the first algorithm superior to the second algorithm, while $R^-$ is the opposite. If the difference between $R^+$ and $R^-$ is sufficiently large, we may reject the non-significant hypothesis, indicating that the two methods show significant performance differences. In addition, the P-value can be calculated according to $R^+$ and $R^-$. If the P-value is below the standard significance index of 0.05, then this algorithm has a significant advantage over the compared method. As shown in Table 6, the P-values are substantially below the standard significance level of 0.05. In Tables 7 and 8, most P-values are also below 0.05. In summary, the ADDFS algorithm has significant advantages over the compared methods, and is an efficient and stable feature selection method.

## 5. Conclusion and future work

This paper proposes to extract a relatively comprehensive feature set to identify video traffic. To obtain an effective feature subset, a novel ADDFS method is proposed. Moreover, we design a video traffic collection architecture to

Video traffic identification with novel feature extraction and selection method collect video traffic data from different platforms in a campus network environment and used these data to conduct a set of experiments. Experimental results show that the proposed peak point feature can significantly improve identification performance. The proposed ADDFS can also be considerably applied to the task of video traffic identification.

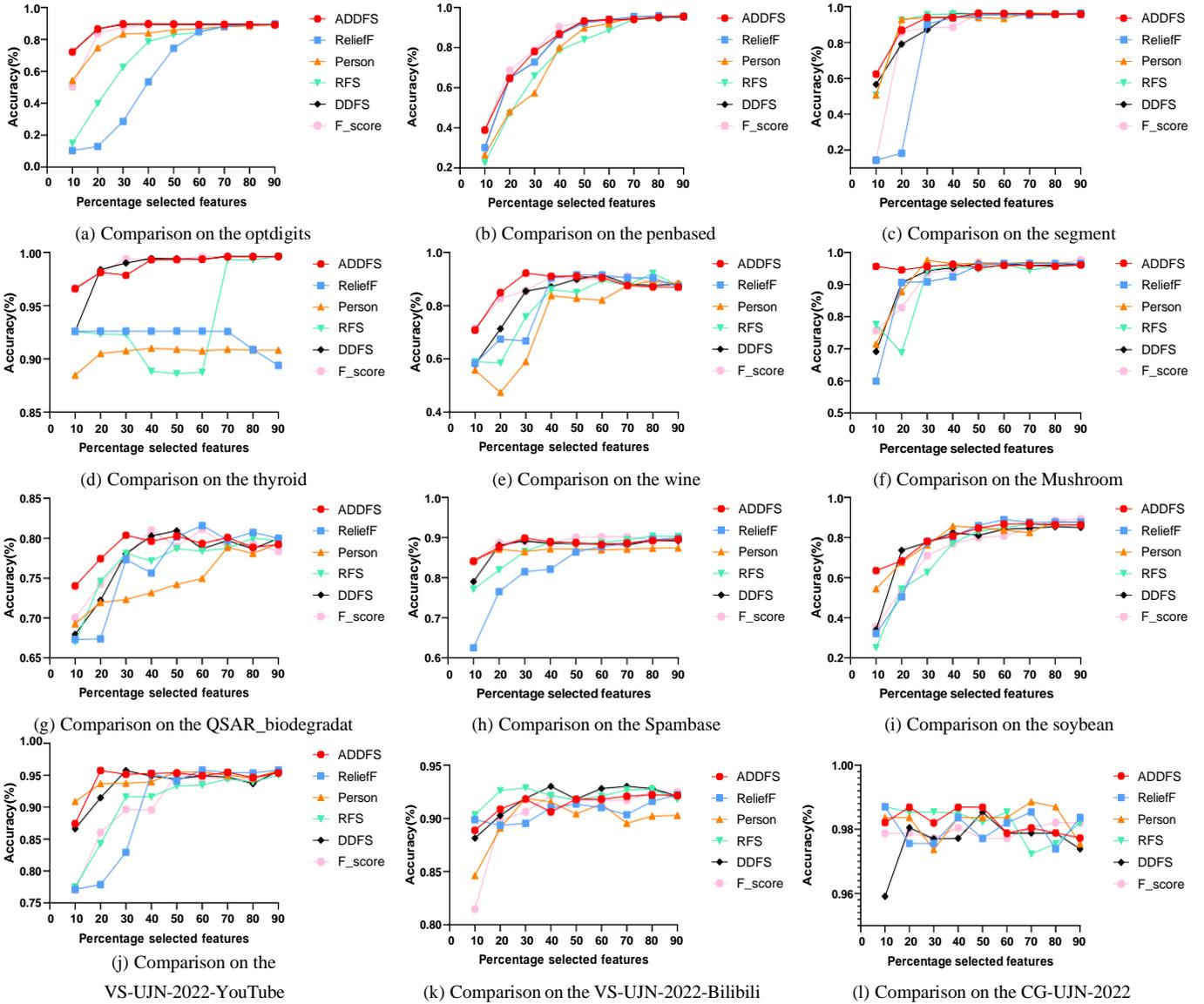

Fig. 11: Results of the compared feature selection methods

Table 7
Wilcoxon signed rank test of ADDFS versus the compared methods-20% of feature set

| Accuracy | | | |
|---|---|---|---|
| ADDFS vs. | R+ | R− | P−value |
| Relief | 78.0 | 0 | 0.0019 |
| Person | 71.0 | 7.0 | 0.0108 |
| RFS | 74.0 | 4.0 | 0.0053 |
| DDFS | 57.0 | 21.0 | 0.1467 |
| F-score | 66.0 | 12.0 | 0.0309 |

Table 8
Wilcoxon signed rank test of ADDFS versus the compared methods-30% of feature set

| Accuracy | | | |
|---|---|---|---|
| ADDFS vs. | R+ | R− | P−value |
| Relief | 78.0 | 0 | 0.0019 |
| Person | 71.0 | 7.0 | 0.0107 |
| RFS | 71.0 | 7.0 | 0.0107 |
| DDFS | 61.0 | 17.0 | 0.0775 |
| F-score | 72.0 | 6.0 | 0.0085 |

Nevertheless, there is also some future research to conduct. First, video traffic data set should be further expanded to include more video scenes and cloud gaming video traffic.

**Table A1**
The description of the extracted features

| Feature Name | Description |
| --- | --- |
| UIAT_mean,min,max,std | Mean, Minimum, Maximum, Std of upstream inter-arrival time interval |
| DIAT_mean,min,max,std | Mean, Minimum, Maximum, Std of downstream inter-arrival time interval |
| IAT_mean,min,max,std | Mean, Minimum, Maximum, Std of all packets inter-arrival time interval |
| UWindow_sum,mean,min,max,std | Sum, Mean, Minimum, Maximum, std of upstream TCP window window sizes |
| DWindow_sum,mean,min,max,std | Sum, Mean, Minimum, Maximum, std of downstream TCP window window sizes |
| Window_sum,mean,min,max,std | Sum, Mean, Minimum, Maximum, std of all packets TCP window window sizes |
| Upnum,Dpnum,pnum | Number of packets for upstream, downstream and all packets |
| Upnum_s,Dpnum_s,pnum_s | The rate of packet number for upstream, downstream and all packets |
| UDpnum_s | The ratio of the packets downstream to the packets upstream |
| FIN,SYN,PSH,ACK_cnt | TCP flag count |
| RST,URG,ECE,CWR_cnt | TCP flag count |
| UPSH,UURG,DPSH,DURG_cnt | Upstream and downstream PSH and URG count |
| Uhdr,Dhdr,hdr | Sum of packet header length for upstream, downstream and all packets |
| UhdrR,DhdrR,hdrR | The ratio of the packet header length sum to the packet payload sum |
| Upay_mean,min,max,std | Mean, Minimum, Maximum, Std of upstream payload |
| Dpay_mean,min,max,std | Mean, Minimum, Maximum, Std of downstream payload |
| pay_mean,min,max,std | Mean, Minimum, Maximum, Std of all packets payload |
| Upayc,Dpayc,payc | Payload peak point count for upstream, downstream and all packets |
| PPP5_mean,min,max,std | Mean, Minimum, Maximum, and Std of PPP every 5 seconds in the first 60 seconds. |
| UBRPP,DBRPP,BRPP | Byte rate peak point count for upstream, downstream and all packets |
| UBRPPSW_mean,min,max,std | Mean, Minimum, Maximum, Std of upstream BRPP size with sliding window |
| DBRPPSW_mean,min,max,std | Mean, Minimum, Maximum, Std of downstream BRPP size with sliding window |
| BRPPSW_mean,min,max,std | Mean, Minimum, Maximum, Std of all packets BRPP size with sliding window |
| BRPPSW_Q1,_Q2,_Q3 | The first, second and third quartile of all packets BRPP size with sliding window |

Second, current research on video traffic analysis is still in the offline analysis stage. In the future, online video traffic identification must be studied.

## Acknowledgment

This research was partially supported by the National Natural Science Foundation of China under Grant No. 61972176, Shandong Provincial Natural Science Foundation, China under Grant No. ZR2021LZH002, Jinan Scientific Research Leader Studio, China under Grant No. 202228114, Shandong Provincial key projects of basic research, China under Grant No. ZR2022ZD01, Shandong Provincial Key R&D Program, China under Grant No. 2021SFGC0401, and Science and Technology Program of University of Jinan (XKY1802)).

## CRediT authorship contribution statement

**Licheng Zhang:** Methodology, Software, Writing – original draft. **Shuaili Liu:** Methodology, Software, Writing – original draft. **Qingsheng Yang:** Data processing. **Zhongfeng Qu:** Formal analysis, Supervision. **Lizhi Peng:** Conceptualization, Formal analysis, Supervision, Writing – review & editing.

## A. My Appendix